\documentclass{nature}

\usepackage{graphicx}
\usepackage{amsmath}
\usepackage{amssymb}
\usepackage{color}
\usepackage{dcolumn}
\usepackage{epsfig}
\usepackage{bm}
\usepackage[urlcolor=blue]{hyperref}
\hypersetup{backref, colorlinks=true, linkcolor=blue, citecolor=blue}

\title{Superconductivity above 120 kelvin in a chain link molecule}

\author{Ren-Shu Wang$^{1,2}$, Yun Gao$^{2}$, Zhong-Bing Huang$^{3}$ \& Xiao-Jia Chen$^1$}

\begin{document}

\maketitle

\begin{affiliations}
 \item Center for High Pressure Science and Technology Advanced Research, Shanghai, 201203, China
 \item School of Materials Science and Engineering, Hubei University, Wuhan 430062, China
 \item Faculty of Physics and Electronic Technology, Hubei University, Wuhan 430062, China
\end{affiliations}

\begin{abstract}
The search for new superconducting compounds with higher critical temperatures $T_{c}^{\prime}$s has long been the very heart of scientific research on superconductivity. It took 75 years for scientists to push the $T_{c}$ above liquid nitrogen boiling temperature\cite{wu} since the discovery of superconductivity. These discoveries have had impacts well beyond the realm of condensed matter physics, with effects ranging from improvements in medical diagnostic tools, such as magnetic resonance imaging, to the development of theories about the interiors of neutron stars. So far, the record high $T_{c}$ of about 130 K at atmosphere pressure was reported in some multilayer Hg(Tl)-Ba-Ca-Cu-O compounds\cite{schi}. Meanwhile, sulfur hydride system\cite{droz} holds the highest $T_{c}$ of around 200 K at high pressure of about 150 GPa. While keeping these records for superconductivity, either the toxicity of these superconductors or the requirement of extreme pressure condition for superconductivity limits their technology applications. Here we show that doping a chain link molecule $-$ $p$-terphenyl by potassium can bring about superconductivity at 123 K at atmosphere pressure, which is comparable to the highest $T_{c}$ in cuprates. The easy processability, light weight, durability of plastics, and environmental friendliness of this kind of new superconductor have great potential for the fine-tuning of electrical properties. This study opens a window for exploring high temperature superconductivity in chain link organic molecules.
\end{abstract}

The staring material is an organic compound only consisting of C and H elements with three phenyl rings connected by single C$-$C bond in para position. It is named $para$-terphenyl or $p$-terphenyl, together with $ortho$-terphenyl and $meta$-terphenyl belonging to the three isomers of terphenyl. As a member of oligophenylenes, $p$-terphenyl is used as a laser dye and a sunscreen ingredient\cite{furu1}, and the longest chain member $p$-sexiphenyl with six phenyl rings can be useful as an active layer in organic light-emitting devices (OLED)\cite{tasc}. These oligophenylenes may serve as model compounds to understand the chain length effect on the physical properties of poly($p$-phenylene). The doping (reduction or oxidation) of this kind of material increases considerably their electrical conductivity\cite{shac}, which contributes poly($p$-phenylene) to an important family of conducting polymers. 

In our experiments, we doped potassium into $p$-terphenyl to induce superconductivity. All experiments and measurements were carried on in HPSTAR laboratories located in Shanghai, China. $p$-Terphenyl (99.5\% purity) was purchased from Sigma-Aldrich and purified by vacuum drying method. High-purity potassium metal (99\%, Sinopharm Chemical Reagent) was cut into small pieces and mixed with purified $p$-terphenyl with a mole ratio of 3:1. The mixtures were then loaded into quartz tubes and sealed under high vacuum (1$\times$10$^{-4}$ Pa). The sample tubes were heated at temperature 443$-$533 K for 24$-$168 hours. After the annealing process, the color of the samples turned to black. The samples were then moved to a glove box with the oxygen and moisture levels less than 1 ppm. For each run of experiment, the sample from the same tube was distributed into several nonmagnetic capsules and sealed by germanium varnish for following magnetization and Raman scattering measurements. Magnetization measurements were performed with a SQUID magnetometer (Quantum Design MPMS3) in the temperature range of 1.8$-$300 K. The Raman scattering spectra were collected at room temperature in an in-house system with Charge Coupled Device and Spectrometer from Princeton Instruments in wavelength of 660 nm and power less than 1 mW to avoid possible damage of samples.

Pristine $p$-terphenyl exhibits the diamagnetic behavior in the temperature range of 1.8$-$300 K characterized by the negative magnetic susceptibility (Extended Data Fig.\,1). When electrons were introduced to $p$-terphenyl through doping potassium, superconductivity and paramagnetism were found in the synthesized samples. The typical paramagnetic behavior in a non-superconducting sample (labeled by 21) was well demonstrated by the excellent fitting of the magnetic susceptibility curve to the Curie-Weiss formula (Extended Data Fig.\,1). A similar Curie-like paramagnetism with surprisingly small Pauli component has been observed to superimpose on a temperature-independent diamagnetic susceptibility in doped poly-$p$-phenylene\cite{peo}. These anomalous spin properties together with high conductivity\cite{shac} and growing optical absorption\cite{crec} provide evidence for a conductivity mechanism including bipolarons that carry no spin in poly-$p$-phenylene\cite{bred}.
 
The observations of high-temperature superconductivity in potassium-doped $p$-terphenyl (labeled by 23) were detailed by the magnetization measurements. Figure 1 shows typical four runs of the $dc$ magnetic susceptibility $\chi$ in the applied magnetic fields of 50, 100, 200, and 300 Oe with field cooling (FC) and zero-field cooling (ZFC) in the temperature range of 100$-$150 K. Both FC and ZFC susceptibilities show a sudden decrease at around 123 K. Below 116 K, $\chi$ exhibits a platform in the FC run, while its decrease is slowed down in the ZFC run. This shape of the magnetization susceptibility curve is consistent with the well$-$defined Meissner effect. This feature together with narrow transition width indicates good superconducting properties of this sample. The superconducting transition shifts systematically downwards to the lower temperatures when higher magnetic fields are applied. The systematic evolution in the temperature range of 100-150 K is summarized in Extended Data Fig.\,2. The superconducting transition at temperatures higher than 120 K in this molecule was unambiguously confirmed from these measurements. The observed low-temperature paramagnetic behavior together with the negative $\chi$ above 200 K indicates that this superconducting sample still contains non-superconducting paramagnetic phase and left pristine (Extended Data Fig.\,3).  Such multiple phases have also been observed in K-doped poly($p$-phenylene) oligomers\cite{pari}. 

The Meissner effect for this 123 K superconductor was further demonstrated from the magnetic field dependence of the magnetization (Fig.\,2). In the superconducting state at a temperature of 100 K, the magnetization $M$ decreases with applying magnetic field and then saturates in the field range of 300-1000 Oe (Fig.\,2$a$). The nice linear behavior up to 130 Oe is the fingerprint for the Meissner phase of a superconductor, yielding a lower critical field $H_{c1}$ of 130 Oe based on the later deviation from the linear $M$ vs $H$ behavior. 

Figure 2$b$ presents the magnetization hysteresis up to 3 Tesla measured at 100 K after subtraction of a diamagnetic background signal. These data provide evidence that our sample is a type-II superconductor with a strong vortex pinning. By further applying magnetic field, the superconducting fraction becomes smaller when more vortices penetrate the superconductor. $M$ is thus gradually reduced when the applied magnetic field becomes larger. The fact that $M$ does not go to zero with the applied field up to 3 Tesla together with the still opening of hysteresis is consistent with the expected high upper critical field for such a 123 K superconductor. 

Raman spectroscopy is an incredibly effective method for identifying the phases and monitoring the process of their formation. Five regions of Raman active modes from the low to high frequencies correspond to the lattice, C$-$C$-$C bending, C$-$H bending, C$-$C stretching, and C$-$H stretching modes. We observed all these modes in pristine $p$-terphenyl. The spectra of potassium-doped $p$-terphenyl are quite different as compared to the pristine material. A comparison to previous studies shows that the observed Raman spectra and features are in good agreement with those in potassium-doped $p$-terphenyl synthesized with liquid ammonia\cite{pere}.

Upon doping potassium into $p$-terphenyl, all lattice modes are significantly suppressed. The C$-$C$-$C bending modes almost resemble the features in the pristine material with the expense of the intensities, indicating that the doped material still maintains the molecular structure. The big difference of the spectra between the pristine and doped samples is in both the C$-$H bending and the C$-$C stretching regions. The modes in the C$-$H stretching region become invisible. 

The out-of-plane bending mode at 771 cm$^{-1}$ and the inter-ring C$-$C stretching modes at 1290 and 1313 cm$^{-1}$ in the doped sample corresponding to the 798 and 1275 cm$^{-1}$ modes in the pristine may be assigned to the polaronic modes, while the C$-$H bending of external rings at 1165 cm$^{-1}$ could be a new polaronic one\cite{dubo,furu2}. The 771 cm$^{-1}$ band can be used as a measure of the length of localization of a polaron\cite{furu2}. The presence of polarons in our samples offers an interpretation of the low-temperature paramagnetic behavior shown in Extended Data Fig.\,3. This turns out true for the observations of superconductivity at low $T_{c}^{\prime}$s of 4.3 and 7.2 K in K-doped $p$-terphenyl with the absence of polaronic bands\cite{wang1}.

The strong bands for the inter-ring C$-$C stretching at 1275 cm$^{-1}$ and the intra-ring C$-$C stretching at 1593 and 1605 cm$^{-1}$ in the pristine behave differently with doping. The former shows upshifts to 1290, 1313, and 1348 cm$^{-1}$, giving the evidence for the formation of polarons and bipolarons, respectively\cite{dubo,furu3}. By contrast, the latter bands merge to the bipolaronic bands centered at 1588 cm$^{-1}$ with the slight downward shifts in wavenumber\cite{dubo}. Here the upshifts mainly reflect the decrease of the C$-$C bonds between rings. The downshifts are the result of the increase of inclined C$-$C bonds within the rings. These effects together drive the structural change of the molecule from the benzenoid to quinoid\cite{bred,dubo}.

A bipolaronic band at 1216 cm$^{-1}$ corresponds to the 1222 cm$^{-1}$ band in the pristine material, arising from a mode concentrated in the inner region of the molecule without any contribution from the terminal rings\cite{ohts}. The bands at 989 and 1474 cm$^{-1}$ with their high-frequency shoulders signal the formation of bipolarons\cite{dubo,furu3}. The presence of 1474 cm$^{-1}$ can be considered as the fingerprint of bipolarons. The relatively high intensity with respect to the background and the growing bands nearby at 1492 and 1517 cm$^{-1}$ indicate that the material is highly doped\cite{simo}.  It is apparent that high concentration of bipolarons is in favor of high $T_{c}$ in synthesized samples\cite{wang2}. The bipolarons may also account for the observed diamagnetic behavior above 200 K in the doped samples (Extended Data Figs.\,1 \& 3).

The 123 K superconducting transition is the highest yet reported for a molecular superconductor compared to the previously reported superconductivity with the record high $T_{c}$ of 38 K in Cs$_{3}$C$_{60}$\cite{gani}. The unique feature of $p$-terphenyl-like molecules is the presence of the chain connecting the phenyl rings. Doping this kind of compounds leads to the formation of polarons and bipolarons. The spinless bipolaron carrying an electric charge $\pm$2$e$ can be thought of as analogous to the Cooper pair in the BCS theory of superconductivity. Whether the chain structure\cite{litt} or bipolaron\cite{chak,alex} could play the key role for such high temperature superconductivity in our sample could be a future attractive topic, the present finding of 123 K superconductivity is very encouraging for the future search of new superconductors in organic molecules connecting the phenyl rings by single C$-$C bond in a chain structure.

\vspace{-5mm}
\subsection{Online Content}
Any additional Methods, Extended Data display items and Source Data are available in the online version of the paper; references unique to these sections appear only in the online paper.

\begin{addendum}

 \item This work was supported by the Natural Science Foundation of China. We thank Hai-Qing Lin and Ho-Kwang Mao for strong support and valuable help.

 \item[Author Contributions] X.J.C., Y.G., and Z.B.H. designed the project. R.S.W. and X.J.C. synthesized the samples and performed the Raman scattering and magnetization measurements. All authors analysed the data and discussed the results. X.J.C. wrote the paper with the inputs of all authors. 

 \item[Author Information] Reprints and permissions information is available at www.nature.com/reprints. The authors declare no competing financial interests. Readers are welcome to comment on the online version of the paper. Correspondence and requests for materials should be addressed to X.J.C. (xjchen@hpstar.ac.cn) or Y.G. (gaoyun@hubu.edu.cn) or Z.B.H. (huangzb@hubu.edu.cn).

 \item[Competing financial interests] The authors declare no competing financial interests.

\end{addendum}

\newpage 

\begin{figure}
\caption{\textbf{$|$ \,Temperature dependence of the dc magnetic susceptibility $\chi$ for potassium-doped $p$-terphenyl.} Four panels show the results measured in the applied magnetic fields of 50, 100, 200, and 300 Oe with field cooling (FC) and zero-field cooling (ZFC) in the temperature range of 100$-$150 K. }
\end{figure}

\begin{figure}
\caption{\textbf{$|$ \,Magnetic field dependence of the magnetization $M$ for potassium-doped $p$-terphenyl at 100 K.} {\bf a}, The magnetic field dependence of the magnetization up to 1000 Oe. The lower critical field $H_{c1}$ is marked by an arrow defined by the deviation from the linear $M$ vs $H$ behaviour. {\bf b}, The magnetization hysteresis with scanning magnetic field along two opposite directions up to 3 Tesla after subtraction of a diamagnetic background signal.}
\end{figure}

\begin{figure}
\caption{\textbf{$|$ \,Raman scattering spectra of pristine $p$-terphenyl and potassium-doped $p$-terphenyl collected at room temperature.} Upper left presents the molecular structure of $p$-terphenyl. The sticks in the upper horizontal axis give the peak positions of the vibrational modes in pristine material. }
\end{figure}

\begin{figure}
\caption{{\bf Extended Data Figure 1 $|$ \,Temperature dependence of the $dc$ magnetic susceptibility $\chi$ for pristine $p$-terphenyl and non-superconducting potassium-doped $p$-terphenyl (labelled by 21).} The data were taken in the zero-field cooling run measured in the applied magnetic fields of 10 and 20 Oe, respectively. The diamagnetic behaviour of the magnetic susceptibility with negative value was observed in pristine $p$-terphenyl over the temperature range of 1.8$-$300 K and doped sample at high temperatures near 300 K. This is probably due to the electronic core contribution. The paramagnetic character at low temperatures can be written as $\chi=\chi_{0}+\chi_{Pauli}+\chi_{Curie}$. A Curie-like contribution ($\chi_{Curie}=C/T$) to the magnetic susceptibility superimposed on a small and temperature-independent contribution has been generally observed in conducting polymers. The fitting with the Curie expression, taken in the temperature range of 1.8$-$200 K, indicates the paramagnetic character. These results imply that potassium-doped $p$-terphenyl may still contain some amount of pristine $p$-terphenyl. }
\end{figure}

\begin{figure}
\caption{{\bf Extended Data Figure 2 $|$ \,Temperature dependence of the $dc$ magnetic susceptibility $\chi$ for potassium-doped $p$-terphenyl in the temperature range of 100-150 K}. The measurements were taken at various magnetic fields up to 1.0 Tesla in the zero-field-cooling (ZFC) run. The high-temperature diamagnetic contribution to the data has been subtracted. }
\end{figure}

\begin{figure}
\caption{{\bf Extended Data Figure 3 $|$ \,Temperature dependence of the $dc$ magnetic susceptibility $\chi$ for potassium-doped $p$-terphenyl.} The data were taken in the applied magnetic field of 200 Oe by the field cooling (red colour) and zero-field cooling (blue colour) runs in the temperature range of 1.8-300 K.  A sharp superconducting transition at 122 K in both FC and ZFC runs was clearly observed. In the normal state, $\chi$ decreases slightly with a sign change from positive to negative around 200 K, indicating that the diamagnetic contribution from pristine $p$-terphenyl is maintained in the synthesized sample. The gradual upturn of $\chi$ in the low-temperature superconducting state comes from either the paramagnetic component of non-superconducting potassium-doped $p$-terphenyl or the polaronic contribution. Thus, the synthesized sample contains both polaronic and bipolaronic contributions. }
\end{figure}

\newpage
\vspace{8cm}
\begin{center}
\includegraphics[width=\columnwidth]{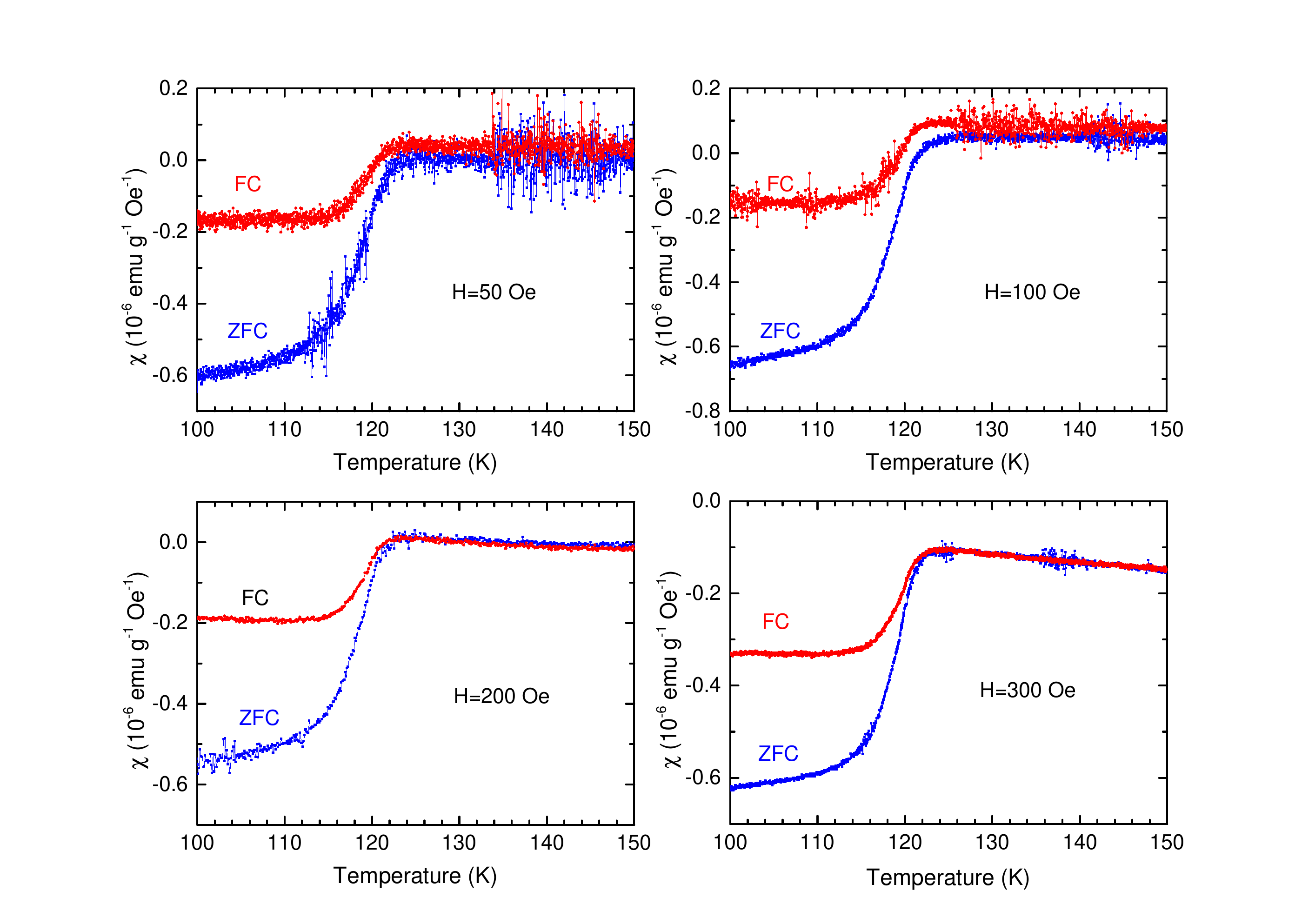}
{\item Figure 1.}
\end{center}

\newpage
%\vspace{8cm}
\begin{center}
\includegraphics[width=0.8\columnwidth]{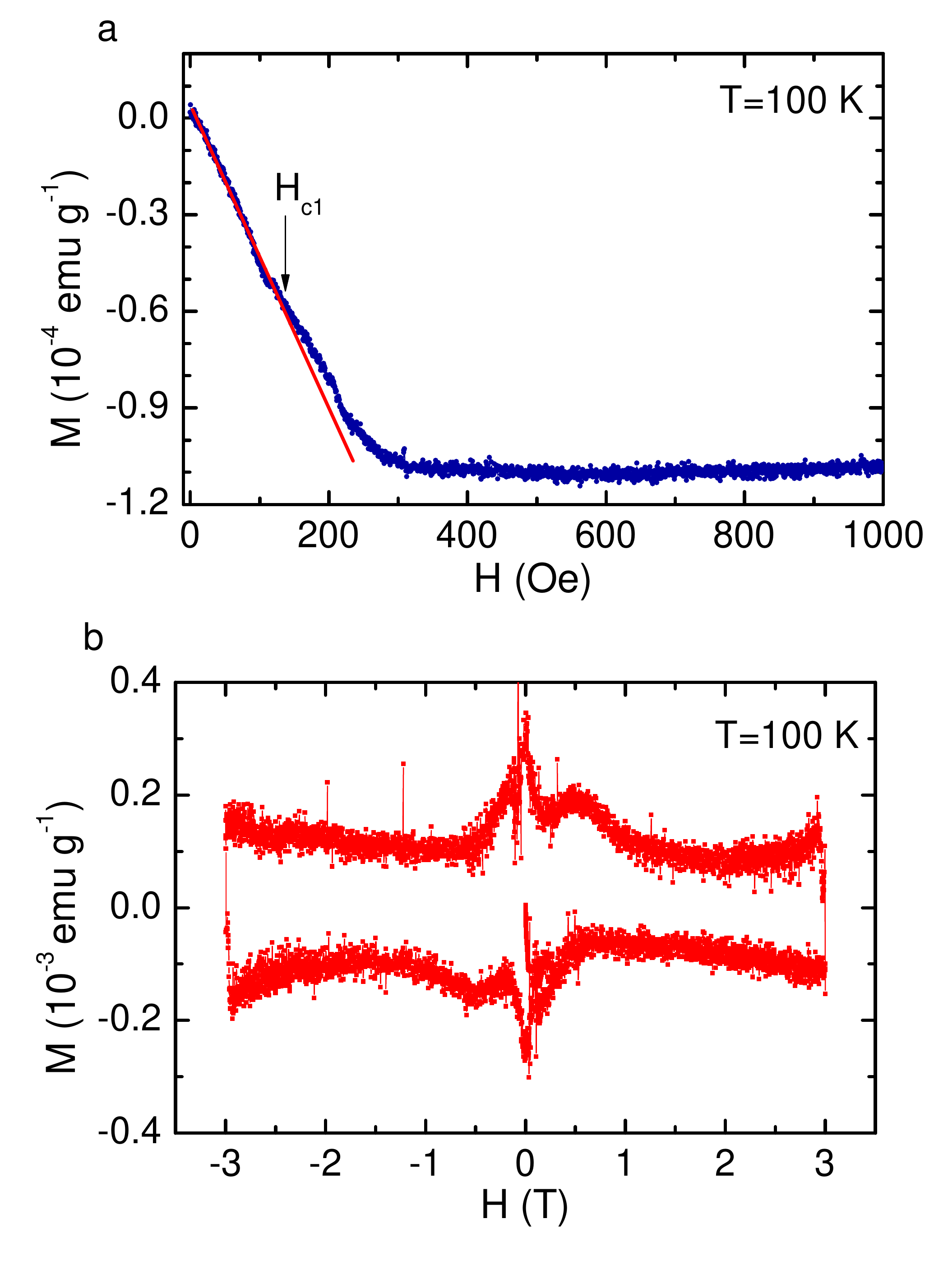}
{\item Figure 2.}
\end{center}

\newpage
\vspace{8cm}
\begin{center}
\includegraphics[width=\columnwidth]{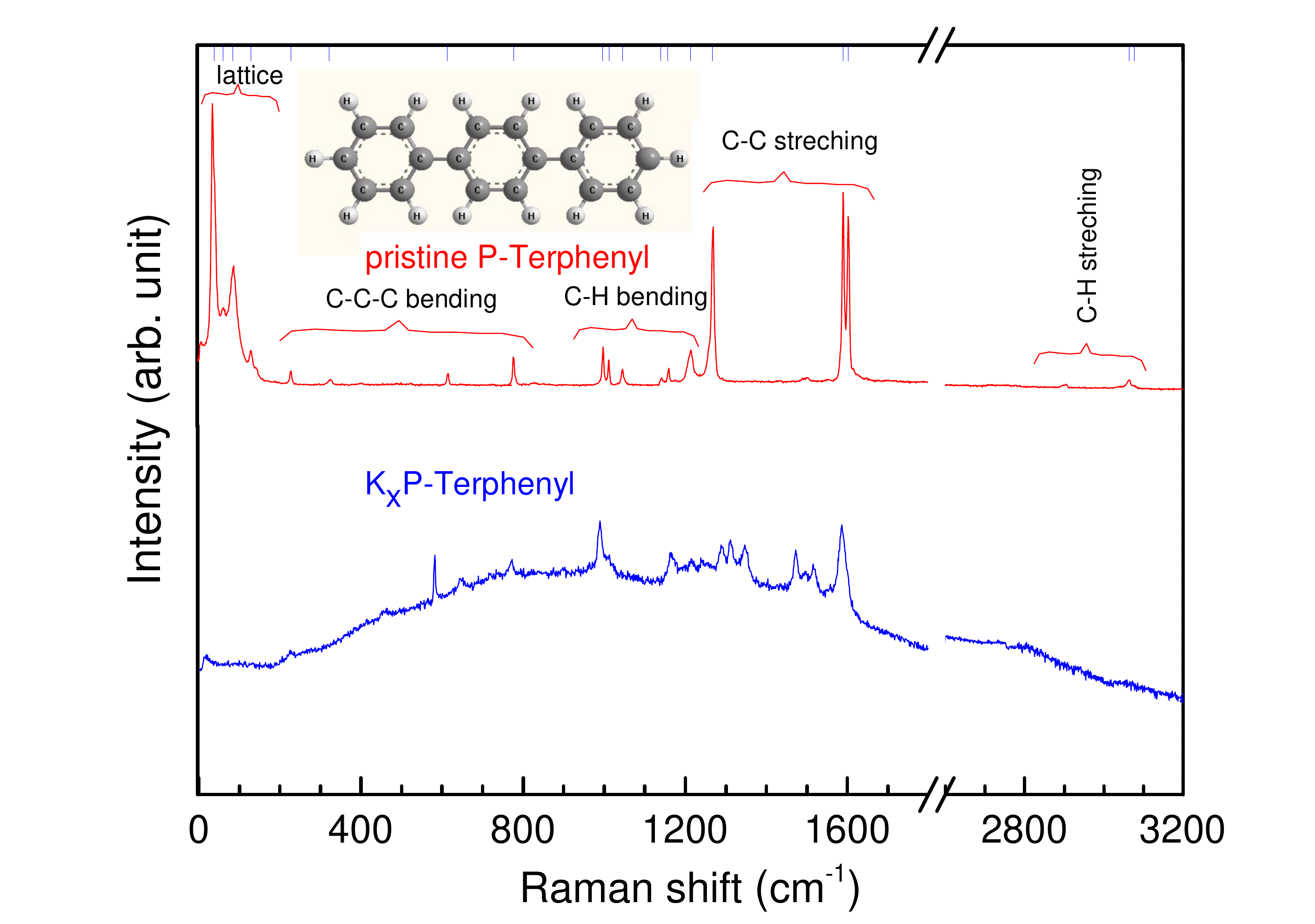}
{\item Figure 3.}
\end{center}

\newpage
\vspace{8cm}
\begin{center}
\includegraphics[width=\columnwidth]{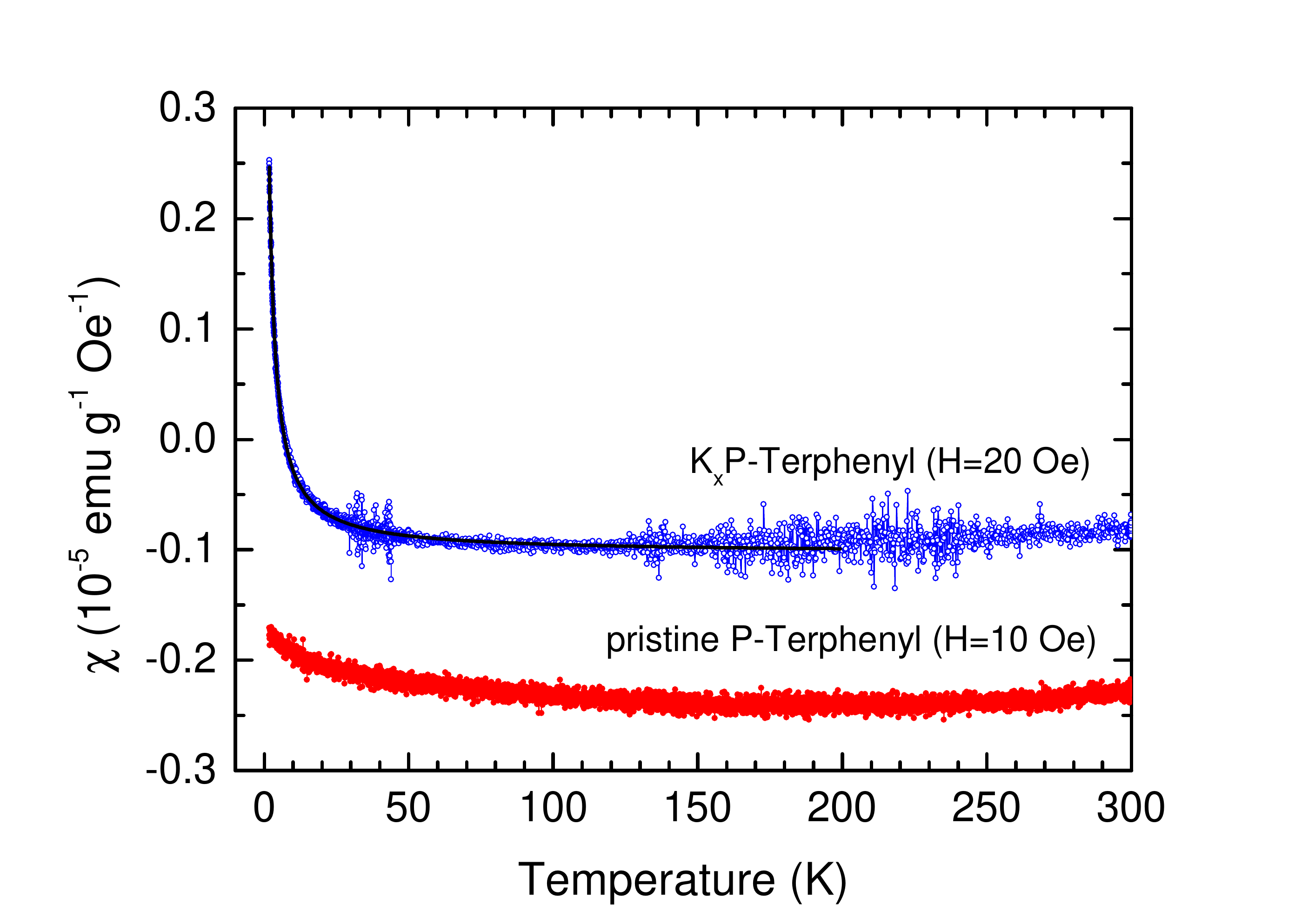}
{\item Extended Data Figure 1.}
\end{center}

\newpage
\vspace{8cm}
\begin{center}
\includegraphics[width=\columnwidth]{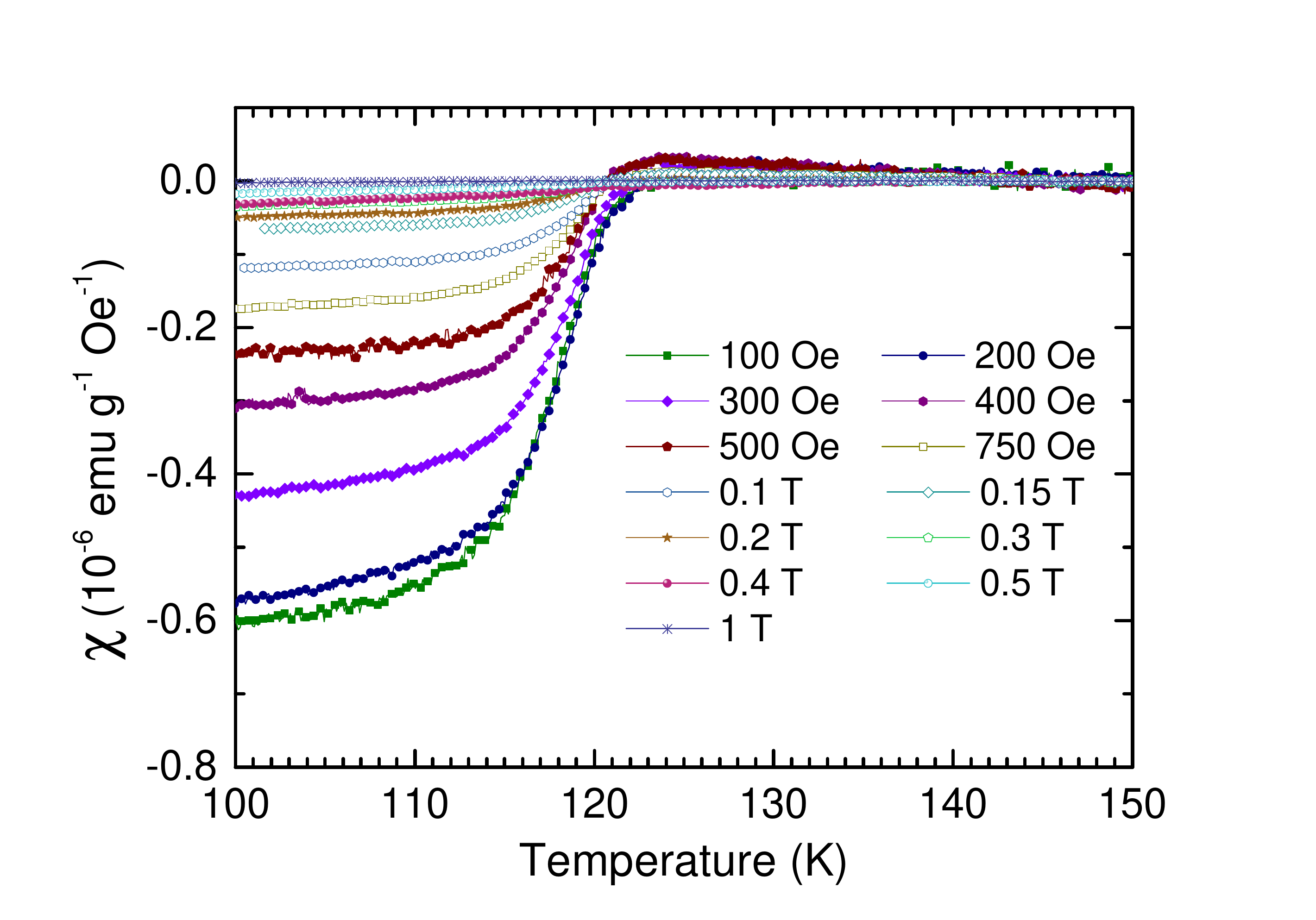}
{\item Extended Data Figure 2.}
\end{center}

\newpage
\vspace{8cm}
\begin{center}
\includegraphics[width=\columnwidth]{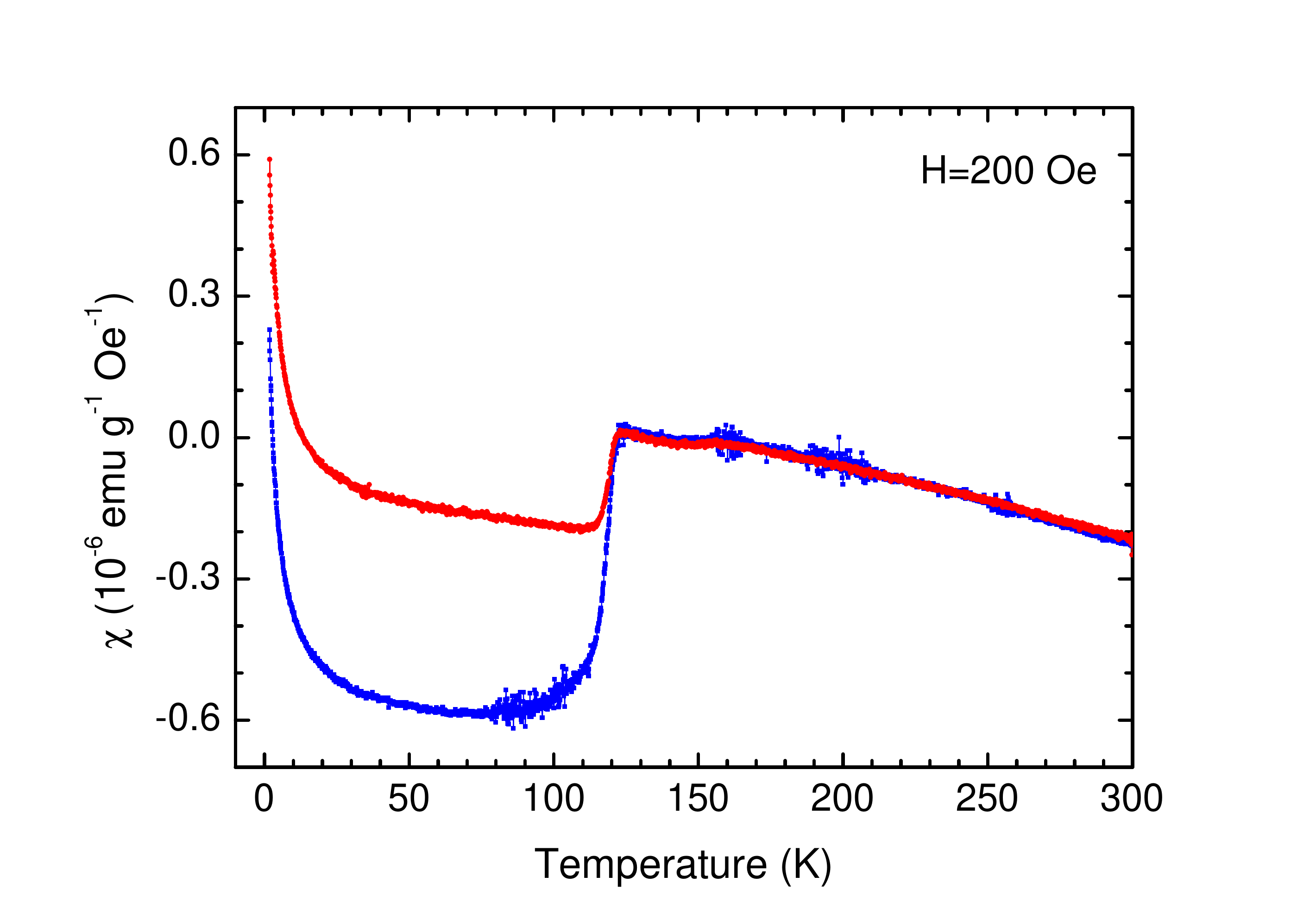}
{\item Extended Data Figure 3.}
\end{center}

\end{document}